\documentclass[]{emulateapj}
\usepackage{natbib}
\usepackage{mathptmx}

\def\griz{$griz$}
\def\gr{$g-r$}
\def\ri{$r-i$}
\def\iz{$i-z$}

\def\OII{[O~II]$\lambda$3727}
\def\OIII{[O~III]$\lambda\lambda$4959,5007}
\def\O5007{[O~III]$\lambda$5007}
\def\Mg2{Mg~II}
\def\Hb{H$\beta$}

\def\Lx{$L_{\rm X}$}
\def\Lxsoft{\Lx$_{(0.5-2~{\rm keV})}$}

\def\Lxtotal{\Lx$_{(0.2-10~{\rm keV})}$}

\slugcomment{Accepted for publication in the Astrophysical Journal,
  November 4, 2004}

\shorttitle{\sc The Nature of Blue Cores in Spheroids}
\shortauthors{\sc Felipe Menanteau et al.}

\begin{document}

\title{The Nature of Blue Cores in Spheroids: a Possible Connection
  with AGN and Star Formation\altaffilmark{\dag}} 

\author{Felipe Menanteau\altaffilmark{1},
Andr\'e~R.~Martel\altaffilmark{1},
Paolo~Tozzi\altaffilmark{2},
Brenda~Frye\altaffilmark{3},
Holland~C. Ford\altaffilmark{1},
Leopoldo~Infante\altaffilmark{4},
Narciso~Ben\'{\i}tez\altaffilmark{1,5},
Gaspar Galaz\altaffilmark{4},
Daniel Coe\altaffilmark{1},
Garth D. Illingworth\altaffilmark{6},
George F. Hartig\altaffilmark{7} and
Marc Clampin\altaffilmark{8}}

\altaffiltext{1}{Department of Physics and Astronomy, Johns Hopkins
University, 3400 North Charles Street, Baltimore, MD 21218.}
\altaffiltext{2}{INAF Osservatorio Astronomico di Trieste, via Tiepolo
  11, I-34131 Trieste, Italy.}
\altaffiltext{3}{Department of Astrophysical Sciences, Peyton Hall - Ivy Lane
Princeton, NJ 08544.}
\altaffiltext{4}{Departmento de Astronom\'{\i}a y Astrof\'{\i}sica,
Pontificia Universidad Cat\'olica de Chile, Casilla 306, Santiago
22, Chile.}
\altaffiltext{5}{Instituto de Astrof\'\i sica de Andaluc\'\i a (CSIC),
  C/Camino Bajo de Hu\'etor, 24, Granada, 18008, Spain.}
\altaffiltext{\dag}{Based on observations obtained at Las Campanas Observatory}
\altaffiltext{6}{UCO/Lick Observatory, University of California, Santa
Cruz, CA 95064.}
\altaffiltext{7}{STScI, 3700 San Martin Drive, Baltimore, MD 21218.}
\altaffiltext{8}{NASA Goddard Space Flight Center, Laboratory for
Astronomy and Solar Physics, Greenbelt, MD 20771.}


\begin{abstract}

We investigate the physical nature of blue cores in early-type
galaxies through the first multi-wavelength analysis of a
serendipitously discovered field blue-nucleated spheroid in the
background of the deep ACS/WFC \griz\ multicolor observations of the
cluster Abell 1689. The resolved \gr, \ri\ and \iz\ color maps reveal
a prominent blue core identifying this galaxy as a ``typical'' case
study, exhibiting variations of $0.5-1.0$~mag in color between the
center and the outer regions, opposite to the expectations of reddened
metallicity induced gradients in passively evolved ellipticals. From a
Magellan-Clay spectrum we secure the galaxy redshift at $z=0.624$. We
find a strong X-ray source coincident with the spheroid
galaxy. Spectral features and a high X-ray luminosity indicate the
presence of an AGN in the galaxy. However, a comparison of the X-ray
luminosity to a sample derived from the Chandra Deep Field South
displays \Lx\ to be comparable to Type~I/QSO galaxies while the
optical flux is consistent with a normal star-forming galaxy. We
conclude that the galaxy's non-thermal component dominates at
high-energy wavelengths while we associate the spheroid blue light
with the stellar spectrum of normal star-forming galaxies. We argue
about a probable association between the presence of blue cores in
spheroids and AGN activity.

\end{abstract}


\section{Introduction}

Early-type galaxies have been the focal point for probing one of the
main expectations from hierarchical models of galaxy formation: the
continuous assembly of galaxies with redshift via
mergers. Considerable attention has been devoted to field ellipticals
at intermediate redshifts with high-resolution multicolor data that
can provide confident morphological classification. The Hubble Deep
Fields (HDFs) and more recently abundant Advanced Camera for Surveys
(ACS) deep imaging has provided significant advances. Recent studies
use the evolution of the fundamental plane for field ellipticals
\citep{Treu-etal-02,vanDokkum-Ellis-03} to constrain their evolution
by comparing to cluster ellipticals, and their colors and number
evolution as a function of redshift \citep[e.g.][]{Bell-etal-04}.

A relatively new approach is the use of color inhomogeneities
exploiting the resolved colors from Hubble Space Telescope (HST)
ellipticals to trace recent star formation activity
\citep{Abraham-etal-99,Papovich-etal-03}. The studies' chief discovery
is that $\simeq30\%$ of HST selected spheroids have internal color
variations, that depart from the expectation for passively evolved
ellipticals (\citealt*{Menanteau-etal-01};
\citealt{Menanteau-etal-04}). Moreover, in most cases the color
variations manifest themselves via the presence of blue cores, an
effect of opposite sign to that expected from metallicity gradients on
passively evolved ellipticals.  Until now, blue cores in ellipticals
have been solely associated with star formation attributed to
differences in local potential well which makes star-formation more
efficient in the central region of the galaxy
\citep{Menanteau-Jimenez-01,Friaca-Terlevich-01}. However, the
physical nature of these objects has been elusive as they have been
only detected in extremely deep HST observations (e.g. the HDFs) and
deep spectroscopic observations have been rather limited
\citep[i.e.][]{vanDokkum-Ellis-03} due to the long integration times
required to acquire high-signal spectrum of faint ellipticals.

In this Letter, we investigate the physical origin of the blue cores
in spheroids through the discovery of a blue-nucleated spheroid galaxy
associated with a strong x-ray source. We use a combination of, deep
ground-based spectrum, HST/ACS deep muticolor observations and
archival Chandra X-ray observations to explore the probable link
between AGN activity and the manifestation of blue central light in
spheroids. We adopt a flat cosmology with $h=0.7$, $\Omega_m=0.3$ and
$\Omega_\Lambda=0.7$ throughout.

\section{Multi-wavelength Observations}

\subsection{Optical Imaging of Abell 1689}

The ACS observations of Abell 1689 were taken in June 2002 as part of
the ACS Guaranteed Time Observations (GTO) science program. They
consist of deep exposures of 4, 4, 3 and 7 orbits in the F475W($g$),
F625W($r$), F775W($i$) and F850LP($z$) bands, respectively. The images
were aligned, cosmic-ray rejected and drizzled together into a single
geometrically corrected image using APSIS (ACS Pipeline Science
Investigation Software; \citealt{Blakeslee-etal-03}) at Johns Hopkins
University, leading to total exposures of 9500, 9500, 11800 and 16600
sec in {\em g, r, i} and {\em z}, respectively, and a final pixel
scale of 0\farcs05~pixel$^{-1}$. 
We refer to \cite{Broadhurst-etal-04} for a more detailed
account of the ACS observations and photometry. Initial object
detection, extraction, and integrated photometry were taken from the
output APSIS SExtractor catalogs using the ACS photometric calibration
in AB magnitudes. Ancillary ground-based imaging in the {\em U} band
and near-infrared {\em J, H, K} bands matching the ACS-defined
apertures were taken from the Abell 1689 catalog of Coe et~al. (in
preparation), which include point spread function (PSF) corrected
ground-based magnitudes.

\begin{figure*}
\plotone{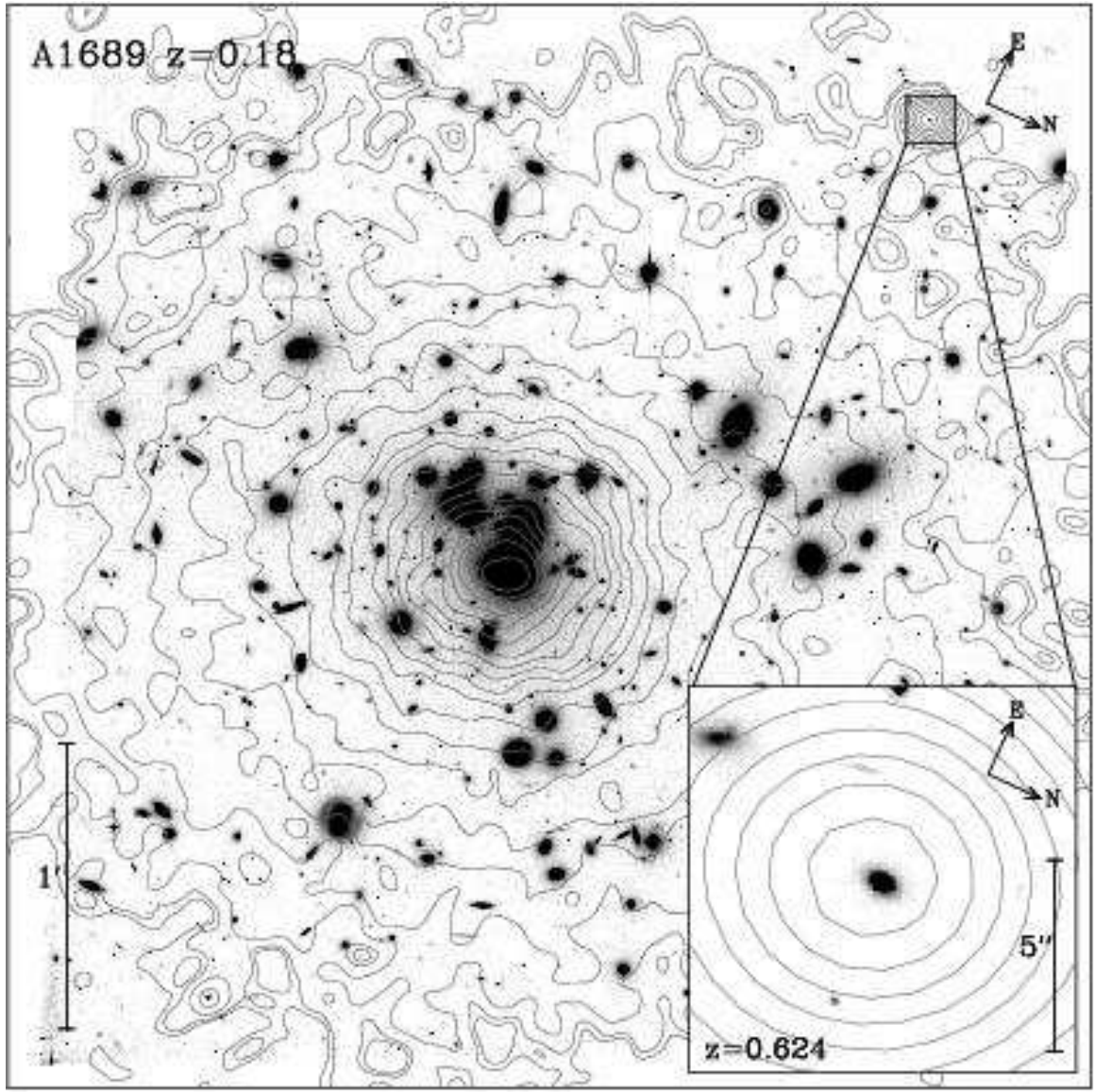}
\caption{The ACS/WFC imaging of Abell 1689 ($z=0.18$) with the
  superimposed Chandra $0.5-7$~keV X-ray contour maps, consisting of
  20 contours between $\sim0.1-14$ counts per pixel displayed with a
  sqrt scale. The inset panel shows an enlargement of the background
  spheroid at $z=0.624$, the strongest field X-ray source in the ACS
  field besides the cluster itself.}
\label{fig:ACS+Chandra}
\end{figure*}

\subsection{Deep LDSS2 Magellan Spectrum}

The spectroscopic information is based on the deep multi-object
observations taken with the Las Campanas Observatory Magellan-Clay
6.5m telescope on May 26, 27 and 28, 2003. We used the low-dispersion
survey spectrograph~2 (LDSS2) as part of a spectroscopic survey of
Abell 1689 (Frye et~al. 2004). Targets for the program were
selected following two criteria: 1) acquiring redshifts for the
multiply-lensed galaxies discovered by \cite{Broadhurst-etal-04}, and
2) securing spectra for high redshift background galaxies. We were
particularly careful to include in all our LDDS2 masks the only
spheroid galaxy with a blue core in the ACS field of Abell 1689.

We used our own IDL reduction software to handle a large number of
spectra efficiently. Optimizing for background-limited data, our aim
was to maximize the signal-to-noise ratio without resampling the data,
so that groups of pixels carrying faint continuum signal have every
chance of being detected as a coherent pattern in the final reduced
image (see \citealt{Frye-etal-02} for details).  The final co-added,
flux-calibrated spectrum of the galaxy totaled 10h through a 1\farcs0
slit in a seeing of $0\farcs6-0\farcs8$ using the Med/Blue grism at
300 lines/mm. Its wavelength dispersion is 5\AA~pixel$^{-1}$,
determined from unblended sky lines, and its spectral coverage,
$\simeq4000-8500$~\AA.

\begin{deluxetable*}{cccccccccccccccc}[b]
\tabletypesize{\scriptsize}
\tablecaption{Multiwavelength Information}
\tablewidth{0pt}
\tablehead{
\multicolumn{2}{c}{(J2000)} & \multicolumn{8}{c}{AB magnitude} & \multicolumn{3}{c}{EW(\AA)} &  \multicolumn{2}{c}{L (erg~s$^{-1}$)} \\ 
\colhead{RA} & 
\colhead{DEC} &
\colhead{U}    &
\colhead{F475W} &
\colhead{F625W} &
\colhead{F775W} &
\colhead{F850LP} &
\colhead{J} &
\colhead{H} &
\colhead{K} &
\colhead{[O II]} &
\colhead{[O III]} &
\colhead{\Hb} &
\colhead{\Lx$_{(0.5-2~{\rm keV})}$} &
\colhead{\Lx$_{(0.5-10~{\rm keV})}$}
}
\startdata
13:11:37.69 & -01:19:49.8 & 21.60 & 22.07 & 21.92 & 21.39 & 21.27 & 20.69 & 20.44 & 20.04 & 23.5 & 70.7 & 9.68 & 2.95$\times10^{43}$ & 1.04$\times10^{44}$ \cr
\enddata
\label{tab:data}
\end{deluxetable*}

\subsection{Chandra X-ray Observations}

We investigated the AGN nature of our blue core spheroid from X-ray
archival data. We used the Abell 1689 observations from the Chandra
X-ray Observatory using ACIS-I in FAINT mode in two exposures of about
10.7 ks and 10.3 ks for a total of 21 ks after data reduction with the
CIAO software. We smoothed out the X-ray image of pixel size
$0.984''$, with a $1.5\sigma$ Gaussian filter and superimposed its
flux contours over the ACS image.  From Fig.~\ref{fig:ACS+Chandra} we
see two main X-ray sources within the ACS coverage of Abell 1689, the
strongest coming from the central cD galaxy in Abell 1689
\citep{A1689-xray} and on the upper right of the ACS image a strong
X-ray source which we positively associate with the blue core spheroid
in our study.

\medskip
We also considered the possibility of radio sources being associated
with the object. Based on radio wavelength archival data from the VLA
FIRST radio survey \citep{FIRST}, we found no radio counterpart
associated with the blue core galaxy at the flux limits of the survey
(0.97~mJy/beam).

\section{Analysis}
\begin{figure*}
\plotone{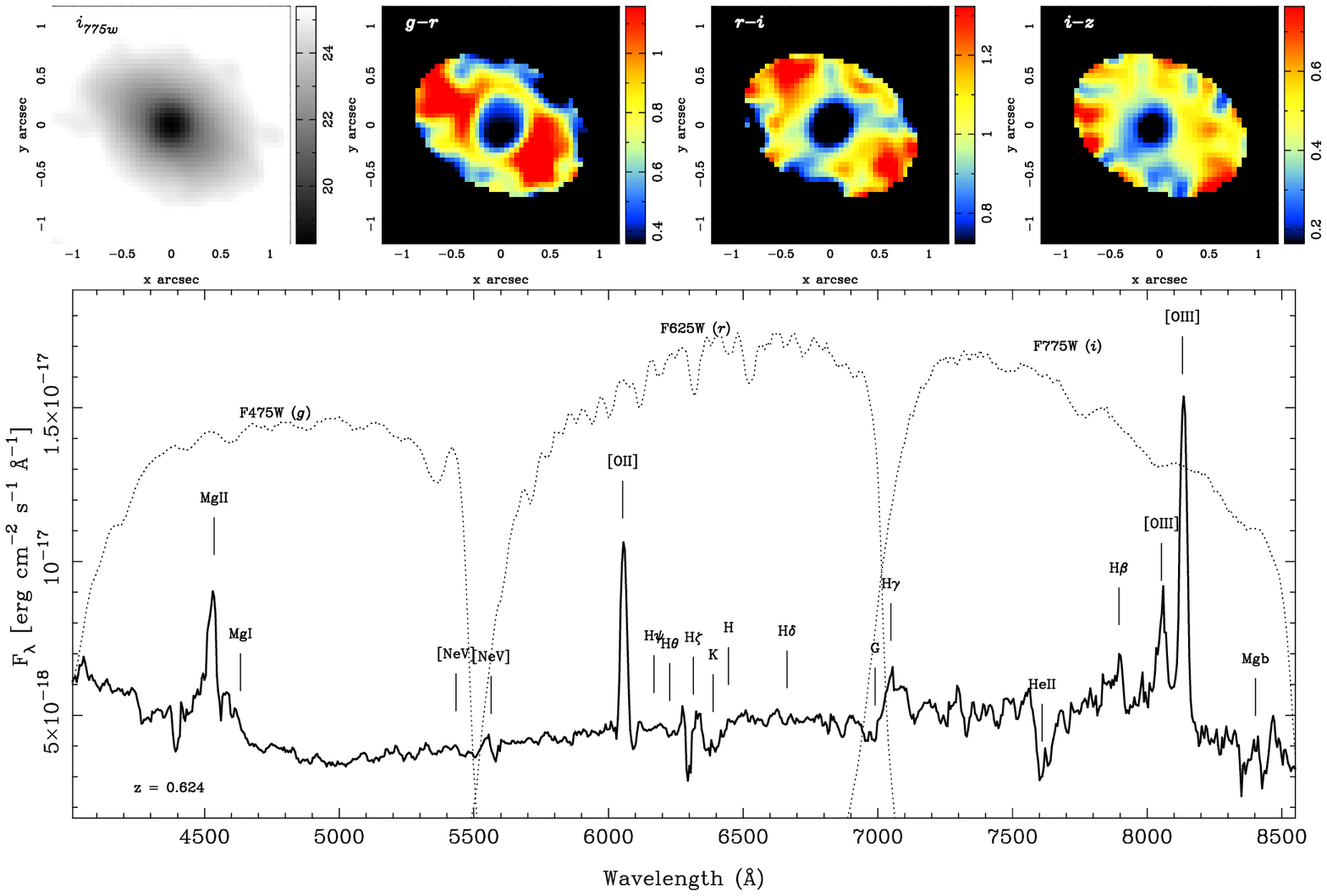}
\caption{Upper panel: the $i$-band surface brightness and the \gr,
  \ri\ and \iz\ color maps for the blue core spheroid from the ACS
  imaging. The figure is color coded to its real color in AB
  magnitudes. Lower panel: the LDSS2 observed spectrum for the galaxy
  at $z=0.624$ is shown as a solid line. The corresponding ACS filter
  bandpasses are shown in arbitrary units as dotted lines. Prominent
  spectral lines are labeled.}
\label{fig:composed}
\end{figure*}
To search for color variations in spheroids, we examine galaxies in
the ACS field of Abell 1689 via the construction of \gr, \ri, and \iz\
color maps for all 742 galaxies with $i_{775w}<24.5$. Although the
effective area for this search is rather limited due to the presence
of the cluster itself and the gravitational lens magnification, we
successfully located one background spheroidal in one of the corners
of the ACS field-of-view with a strong central blue spot similarly to
those reported previously (see upper panel
Fig.~\ref{fig:composed}). This is the only spheroid with a blue core
in the field, including all the objects associated with the
cluster. To study the color maps and galaxy profiles, we first removed
the dependence of the PSF structure on wavelength by deconvolving the
galaxy image in each bandpass with an appropriate PSF using the IRAF
{\em lucy} task with flux conservation. The PSFs were derived from
several observations of a well-exposed star in ACS calibration
programs. The restored images were then convolved identically with the
F625W PSF, thus removing any color dependence due to the PSF shape. We
used this images to create the galaxy resolved color maps. These show
a prominent blue core with color differences of $\simeq1.0$~mag in
\gr, $\simeq0.8$~mag in \ri\ and $\simeq0.9$ in \iz.

We verified if the galaxy possesses an unresolved nucleus, a common
signature of nuclear activity, by comparing its nuclear profile in
each filter with that of an observed PSF. The galaxy radial profiles
were extracted with the IRAF {\em radprof} task with ``background''
subtraction defined in an annulus of radius 6~pixels ($0\farcs3$) and
width 2~pixels, located inside the host galaxy. The resulting galactic
profiles were then compared with those of the PSF star. A positive
point source identification was ascertained in all four filters out to
a radius of 5~pixels ($0\farcs25$) . The match of the galaxy's nuclear
profile with the PSF is best in F475W filter, where the host galaxy
contamination is the least. The F625W profiles exhibit the largest
deviation possibly from contamination by \OII\ line emission (see
Fig.~\ref{fig:composed}).

The spectrum (Fig.~\ref{fig:composed}) shows prominent emission lines
that indicate the presence of an active nucleus and also suggest star
formation activity. The galaxy contains the distinctive \Mg2, \Hb\ and
\OIII\ nebular lines associated with AGNs. However, the signal is not
high enough to see the broad-line component for the \Mg2\ and \Hb\
permitted lines, and H$\alpha$ lies outside the observed spectral
range. The continuum does not follow a pure power-law, but rather
includes some absorption features such as Mgb.

Additional information confirming the presence of an active nucleus
comes from the X-ray observations.  The galaxy is detected as a point
source in the X-ray with a very high signal and with a total of
$\simeq 130$ net counts in the 0.3--10~keV band, extracted from a
circular region of 8\arcsec. 
From a difference of almost 9 months between the two Chandra
observations, the source seems to be variable with a luminosity
decrease of $\sim 30\%$ at a 2 sigma c.l. in the soft band
($0.5-2$~keV). However, we do not detect any variability in the hard
band. This is expected, since variability is routinely found in X-ray
sources in Chandra surveys \citep{Paolillo-etal-04}.
To correct for possible contamination
from the diffuse emission from the cluster, which is very strong in
the position of the galaxy image, we experimented extracting three
different backgrounds around the galaxy and determined that the
background is not affecting our results significantly.
Due to the low number of net detected counts, we used the
\cite{Cash-79} statistics to fit the spectrum of the galaxy with XSPEC
(V11.3) 
the energy range $0.6-8$~keV, to avoid calibration uncertainties at
energies lower than 0.6~keV.  Our model is a simple power-law plus a
Galactic absorption and an intrinsic absorption.  We find a spectral
slope of $\Gamma=1.5 \pm 0.2$ and an upper limit to the intrinsic
absorption of $2\times 10^{21}$ cm$^{-2}$.  The redshift is frozen to
$z=0.624$ and the Galactic absorption to $N_H = 1.82 \times 10^{20}$
cm$^{-2}$. The luminosity in the rest-frame $0.5-2$~keV and
$0.5-10$~keV band is $L_{\rm X} = 2.95 \times 10^{43}$ erg s$^{-1}$
and $L_{\rm X} = 1.04 \times 10^{44}$ erg s$^{-1}$ respectively (see
Table~\ref{tab:data}). Both the absence of significant absorption and
the high luminosity point towards a Type~I AGN. 

We also notice a residual around 4 keV, which is the observing frame
energy expected for a possible Fe line complex at the rest--frame
energy of $\sim6.4$~keV. Therefore, we repeated the fits by adding a
simple Gaussian line and leaving free the energy, the width and
normalization of the line. We obtain the best fit values: $E_{line} =
3.93\pm 0.08$~keV, and equivalent width EW$= 0.9 \pm 0.5$~keV. In
addition, the best fit slope of the power law is $\Gamma = 1.65\pm
0.25$, while the upper limits on the intrinsic absorption are somewhat
larger but consistent with the values obtained without the line. The
decrease in C--statistics with respect to the model without the line
is $\Delta C\sim 5.9$. Considering that we added three free parameters
defining the line, such a decrement correspond to a significance level
of about 90\%. However, since the best--fit energy of the line is what
we expected for the redshifted 6.4 keV K--shell transition from Fe,
the significance of the detected line is more than 2 sigma.
We conclude that we possibly detected a Fe line originated in the
nuclear region, an occurrence to be verified with further observations
already scheduled for A1689 (public release March 2005).

\begin{figure}
\plotone{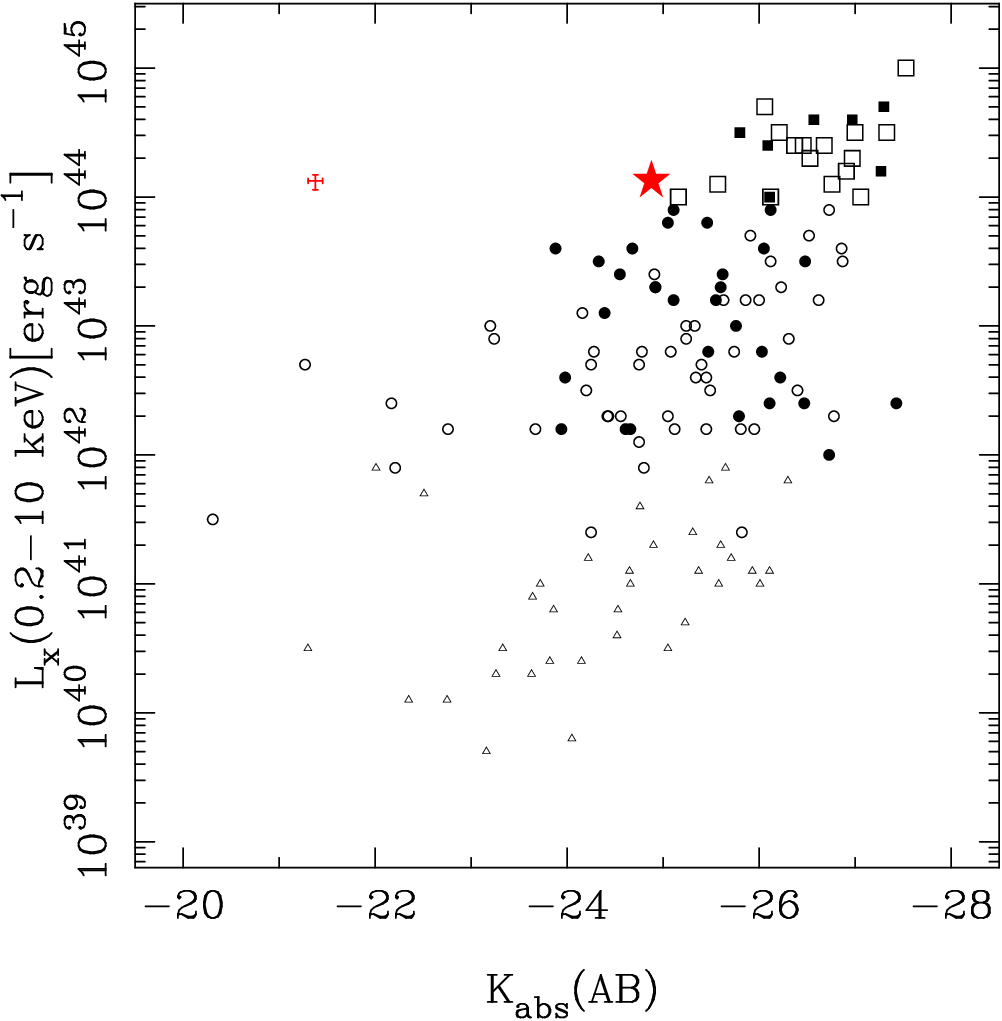}
\plotone{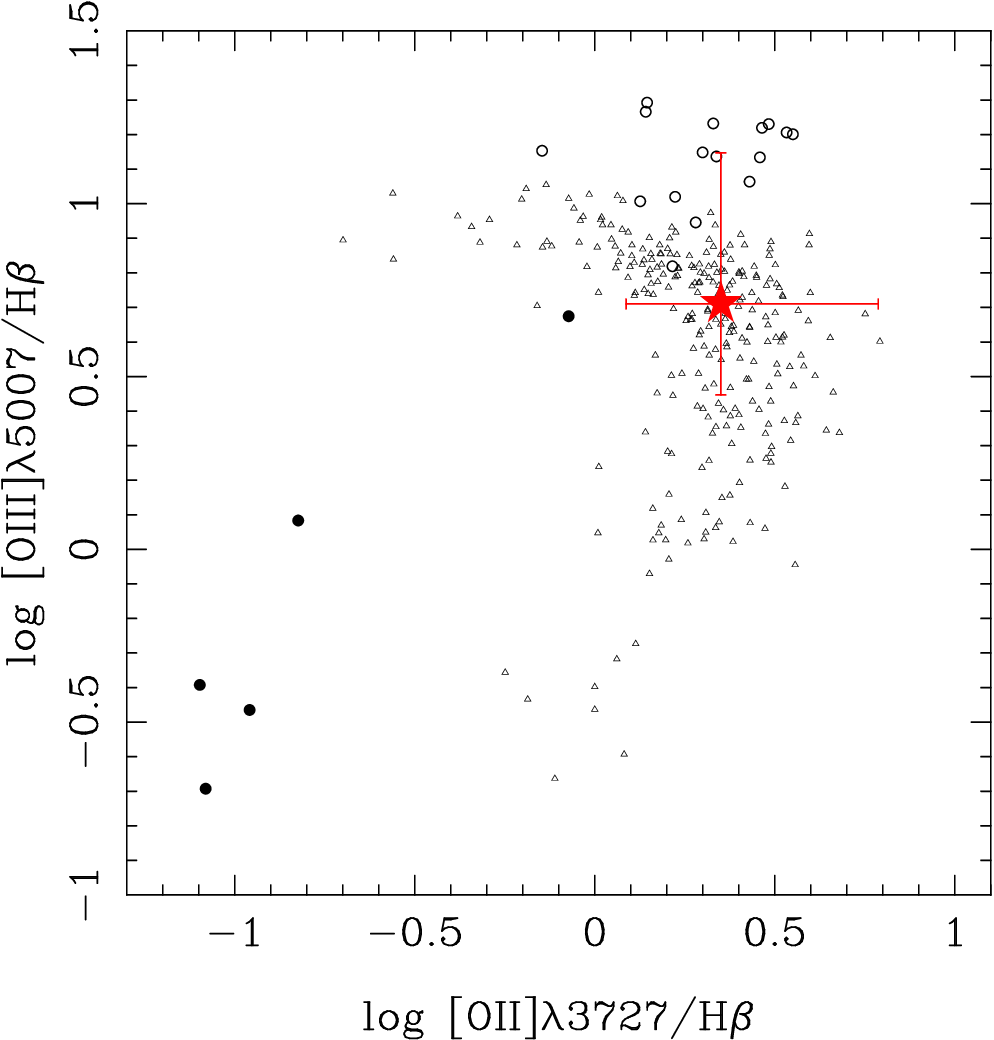}
\caption{Top panel: the X-ray to optical luminosity for the Chandra
  Deep Field South spectroscopic catalog from \cite{Szokoky-etal-04}
  compared to the spheroid galaxy. The error bars for the galaxy have
  been shifted to avoid overlapping with the symbol. Bottom panel: the
  \OII\ to \O5007\ relative to \Hb\ flux from the spectro-photometric
  catalog from \cite{Terlevich-etal-91} compared to the values
  extracted for the spheroid. In both panels the red star represents
  the blue core spheroid; filled and open circles show Type~I and
  Type~II AGNs respectively, squares represent QSO galaxies and open
  triangles normal and HII galaxies.}
\label{fig:OII-Lx}
\end{figure}

\section{Discussion}

We examine the relationship between the X-ray and optical luminosities
of the galaxy and its connection with galaxy types. We compare the
galaxy $L_{X(0.2-10)~{ \rm keV}}$ luminosity and $K$-band absolute AB
magnitude with the galaxies in the \cite{Szokoky-etal-04} Chandra Deep
Field South (CDF-S) catalog of X-ray sources. The $K$ absolute
magnitude is computed using the $k$-correction derived from the Type~I
spectral energy distribution (SED) of the Chatzichristou (private
communication) library of SEDs. In Fig.~\ref{fig:OII-Lx} (upper
panel), we compare \Lx\ and $K$ for the galaxy (solid star) and the
CDF-S sample.  It is interesting to note that the galaxy has an X-ray
luminosity in the same range as low luminosity QSOs (squares) and
Type~I AGNs (filled circles). However, the galaxy's $K$ flux is more
similar to that of Type~II AGNs (open circles) or normal galaxies
(triangles). This suggests that the galaxy's non-thermal component is
very strong and dominant for high-energy wavelengths while there is a
star-forming stellar component that contributes most of the light at
optical wavelengths.

We also investigate the relation between the line flux ratios
\OII$/$\Hb\ and \O5007$/$\Hb. This diagnostic diagram has been used for
the spectral classification of emission-line galaxies
\cite[see][]{Tresse-etal-96}. In Fig.~\ref{fig:OII-Lx} (lower panel),
we compare the line fluxes between the spheroid galaxy (solid star)
and the values for different galaxy types taken from the
spectrophotometric catalog of galaxies of \cite{Terlevich-etal-91}. We
find that the galaxy has \OII$/$\Hb\ values in the same range as
Type~II AGNs and HII galaxies and that its \O5007$/$\Hb\ flux is more
similar to HII and normal galaxies. Moreover, the typical
\O5007$/$\Hb\ and \OII$/$\Hb\ values of Type~I AGNs (solid circles)
are significantly lower than the values of the galaxy.  We consider
this as further evidence for the dual nature of the galaxy. Although
the spectrum shows AGN signatures such as MgII and \OIII, the strength
of the \OII\ line suggest the presence of ongoing stellar formation
processes. If we associate the \OII\ flux to stellar activity we can
make a rough estimate of the galaxy star formation rate (SFR) based on
its \OII\ luminosity, $L[{\rm O~II}]$, using the prescription of
\cite{Kennicutt-98}. Whereas the rates derived from \OII\ are less
precise than those from H$\alpha$ due to uncertainties in the assumed
H$\alpha$ extinction, it is still possible to obtain a reasonable
estimate for the galaxy SFR. We compute $L[{\rm O~II}] = 2.79\times
10^{41}$~ergs~s$^{-1}$ which yields a
SFR~$=3.90\pm1.12$~M$_{\sun}$yr$^{-1}$. This represents a modest rate,
lower than in massive starburst galaxies
\citep{Ranalli-etal-03,2004MNRAS.347L..57G}. However it is consistent with normal disk
galaxies and SFR computed for spheroids with similar blue cores, based
purely on models \citep{Menanteau-etal-01, Menanteau-Jimenez-01}.

The \Lxsoft\ has been employed as a SFR indicator
\citep[see][]{Ranalli-etal-03,Cohen-03}, because of its link to X-ray
binaries, \citep{Persic-etal-04}, young supernova remnants and
galactic winds associated to star-forming galaxies. Additionally
\cite{2004MNRAS.347L..57G} have derived the relation between \Lxtotal
and SFR, consistent with \cite{Ranalli-etal-03} except for low SFR
\citep{2003MNRAS.339..793G}. Could the observed spheroid \Lx\ be
solely associated with star formation? It is hard to rule out that a
fraction of it is not related, however the 2--10~keV hard band
luminosity \Lx$=7.5\times10^{43}$~erg~s$^{-1}$, would suggest an
extremely high SFR ($>10^4$~M$_{\sun}$yr$^{-1}$) in contradiction with
the optical data suggestive of low star-formation activity. There is
also a tight correlation between $L(1.4~{\rm Ghz})$ radio luminosity
and \Lx\ for starburst galaxies \citep{Cohen-03}. However, we report
no signal for the galaxy at radio wavelengths (i.e. \S~2).

\section{Conclusion}

Based on the galaxy's spectral features and its X-ray luminosity, we
have determined the presence of an active non-thermal component in the
blue spheroid galaxy identified in the background field of Abell~1689.
Star formation and AGNs have been suggested to be closely
inter-connected \citep{Levenson-etal-01}. However, they have been
related mostly with Seyfert~2 galaxies which are heavily absorbed in
their hard X-ray emission ---not the case for this galaxy. Moreover, the
fact that we observe such a prominent blue core in the ACS images
suggests that the central region is not particularly affected by dust
obscuration, nor that it contains a dusty enshrouded starburst. On the
other hand most present-day galaxies harbor supermassive black holes,
which may have an important role in the formation of ellipticals and
bulges \citep{Merrit-Ferrarese-01}. It is tantalizing to relate the
galaxy central blue light with the rapid mass infall into a central
black hole which might trigger star formation, and the subsequent
fueling of gas into the AGN which might be the phenomenon we are
observing.

The advent of wide areas with deep publicly available HST/ACS
multicolor imaging and Chandra X-ray observations will make it
possible to examine whether the association between AGNs and blue core
spheroids is a common feature or just a rare occurrence. If they prove
to be a constant feature in blue core spheroids, these might represent
a new galaxy subclass and may provide evidence for a more delayed
formation scenario for early-type galaxies.

\acknowledgments

We thank Ann Hornschemeier for useful conversations on the
subject. ACS was developed under NASA contract NAS 5-32865, and this
research is supported by NASA grant NAG5-7697. LI and GG acknowledge
support from \mbox{FONDAP} "Center for Astrophysics".

\end{document}